\documentclass[11pt]{article}

\usepackage[margin=1in]{geometry}

\usepackage{palatino}

\usepackage[
  colorlinks,
  linkcolor = blue,
  citecolor = blue,
  urlcolor = blue]{hyperref}
\usepackage{amsmath,amssymb,amsthm}
\usepackage{graphicx}

\newcommand{\R}{\mathbb{R}}

\newcommand{\C}{\mathbb{C}}

\newcommand{\ket}[1]{|#1\rangle}
\newcommand{\bra}[1]{\langle#1|}

\newcommand{\abs}[1]{\lvert #1 \rvert}

\newcommand{\ie}{\textit{i.e.}}

\newcommand{\be}{\begin{equation}}
\newcommand{\ee}{\end{equation}}
\def\<{\langle}
\def\>{\rangle}

\DeclareMathOperator{\pr}{\Pr}

\DeclareMathOperator{\tr}{Tr}

\DeclareMathOperator{\Pos}{Pos}

\newcommand{\Thm}[1]{\hyperref[thm:#1]{Theorem~\ref*{thm:#1}}}
\newcommand{\Lem}[1]{\hyperref[lem:#1]{Lemma~\ref*{lem:#1}}}
\newcommand{\Cor}[1]{\hyperref[cor:#1]{Corollary~\ref*{cor:#1}}}
\newcommand{\Def}[1]{\hyperref[def:#1]{Definition~\ref*{def:#1}}}
\newcommand{\Obs}[1]{\hyperref[obs:#1]{Observation~\ref*{obs:#1}}}

\newcommand{\Sect}[1]{\hyperref[sec:#1]{Section~\ref*{sec:#1}}}

\newcommand{\Fig}[1]{\hyperref[fig:#1]{Figure~\ref*{fig:#1}}}
\newcommand{\Tab}[1]{\hyperref[tab:#1]{Table~\ref*{tab:#1}}}
\newcommand{\EqRef}[1]{\hyperref[eq:#1]{(\ref*{eq:#1})}}
\newcommand{\Eq}[1]{Equation~\hyperref[eq:#1]{(\ref*{eq:#1})}}

\newtheorem*{example*}{Example}

\theoremstyle{definition}

\title{A unified view on Hardy's paradox and the CHSH inequality}
\author{Laura Man\v{c}inska and Stephanie Wehner}

\begin{document}
\maketitle

\begin{abstract}
Bell's inequality fundamentally changed our understanding of quantum mechanics. Bell's insight that
non-local correlations between quantum systems cannot be explained classically can be verified
experimentally, and has numerous applications in modern quantum information. 
Today, the CHSH inequality is probably the most well-known Bell inequality and it has given 
us a wealth of understanding in what differentiates the classical from the quantum world. 
Yet, there are certainly other means of quantifying ``Bell non-locality without inequalities''
such as the famous Hardy's paradox. As such, one may wonder whether these are entirely different approaches to non-locality. 
For this anniversary issue, we unify the perspective of the CHSH inequality and Hardy Paradox
into one family of non-local games which include both as special cases.
\end{abstract}

\section{Introduction}
\label{sec:Intro}

Fifty years ago, Bell's seminal work~\cite{Bell64} dramatically changed our understanding of nature. 
He presented a simple inequality involving measurements on two systems which 
is satisfied in any classical theory, that is, a theory in which the answer to any measurement is already (probabilistically) recorded in the form of local hidden variables and merely revealed during the measurement process.
He then proceeded to show this inequality can be violated if the nature is governed by quantum mechanics.
Bell's inequality 
is deceptively simple, but increasingly accurate 
experimental verification~\cite{Freedman72,Fry76,Aspect81,Aspect82b,Aspect82a,Kwiat95, Pan00, Rowe01, Salart08, Ansmann09,Giustina13,Christensen13} 
shows that such a violation can indeed be observed. This forces us to accept that nature really does
not admit any underlying classical description in terms of local hidden variables (see~\cite{Brunner13} for a survey). 

In recent years, Bell's inequality has risen to prominence in quantum information, enabling,
for example, quantum key distribution~\cite{Bennett84,Ekert91} and the generation of 
certified randomness~\cite{Colbeck06, Pironio10}. Indeed the term \emph{Bell inequality} now
refers to any inequality that places a bound on the strength of correlations
that we can observe between two (or more) spacelike separated classical 
systems. 
As it is commonly done~\cite{Brunner13}, we thereby use the 
magnitude of the violation of a Bell inequality to quantify the amount of non-locality exhibited by quantum mechanical strategies.

Arguably the most famous of Bell inequalities is the 
so called CHSH (Clauser-Horne-Shimony-Holt) inequality \cite{Clauser69}. 
Let us follow the quantum information convention and thereby call the first system Alice and the second Bob. The CHSH inequality involves two measurement settings per party, which we will denote as $A^0$ and $A^1$ for Alice and $B^0$ and $B^1$ for Bob. Each measurement has two outcomes, $+1$ and $-1$. 
Concretely, the CHSH inequality bounds the following linear combination of expectations 
\be
  \<CHSH\>:=\<A^0 B^0\> + \<A^0 B^1\> 
  + \<A^1 B^0\> - \<A^1 B^1\>\ ,
\label{eq:chsh}
\ee
where $\<A^iB^j\>$ is the expected value of the product of the outcomes produced when Alice measures $A^i$ and Bob measures $B^j$. 
Since the measurement outcomes are $\pm 1$, the value of $\<CHSH\>$ cannot exceed $4$. In order to achieve this value, it is necessary (although not sufficient) for Alice and Bob to ensure that their measurement outcomes are perfectly correlated (\ie, Alice's outcome determines Bob's outcome and vice versa). In general, higher values of of the expectation $\<CHSH\>$ require stronger correlation between the measurement outcomes.
The CHSH inequality bounds the strength of classically achievable correlations, $\<CHSH\>_c$ as
\be
  \<CHSH\>_c \le 2.
\ee
Quantum mechanics allows to achieve higher values thus violating the CHSH inequality. Indeed, we obtain $2\sqrt{2}$ if Alice and Bob measure their respective parts of $\ket{\Psi^-}:=\frac{1}{\sqrt{2}}\big(\ket{01}-\ket{10}\big)$ using $A^0 := \sigma_z$,  $A^1 := \sigma_x$ for Alice and $B^0 := -(\sigma_x+ \sigma_z)/\sqrt{2}$,  $B^1 := (\sigma_x-\sigma_z)/\sqrt{2}$ for Bob. In fact, Tsirelson \cite{Tsirelson80} has showed that $2\sqrt{2}$ is the highest expectation that can be achieved with quantum mechanics, \ie,
\be
\max \<CHSH\>_q = 2\sqrt{2}\ ,
\ee
where the maximization is taken over all quantum states and measurements of Alice and Bob.

\begin{figure}[ht]
\centering
  \includegraphics{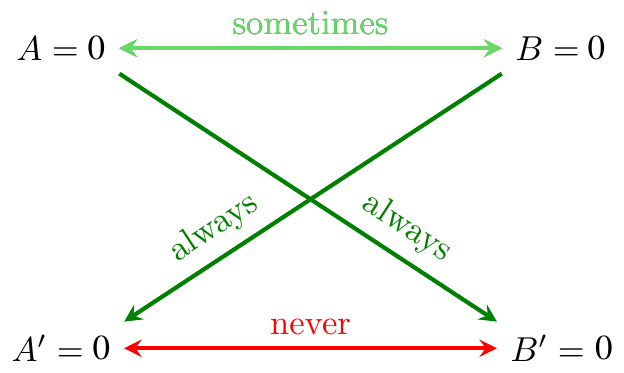}
\caption{Schematic depiction of the probabilities obtained in Hardy's setup. If both parties perform the unprimed measurements, they both obtain zero with positive probability. In contrast, they never both obtain zero when performing the primed measurements. If one of the parties performs the unprimed measurement and obtains zero, then the other party must also obtain zero upon the primed measurement.}
\label{fig:Hardy}
\end{figure}

\paragraph{Hardy's Paradox.}
Of course one may question whether Bell inequalities are the only way to measure the strength of non-local correlations and allow us to distinguish the classical from a quantum world. Indeed, one other measure is Hardy's seminal result known as Hardy's paradox~\cite{Hardy92,Hardy93}, also referred to as ''non-locality without inequalities''.
Just like CHSH it involves two parties each of which have two binary outcome measurement settings.
Yet unlike CHSH and Bell inequalities in general, in Hardy's paradox the difference between classical and quantum worlds lies within the 
\emph{possibility} of occurrence of some type of events rather than a combination of expectation values.
Concretely, Hardy's paradox arises out of the following measurement setup, summarized in Figure~\ref{fig:Hardy}.
Alice and Bob can each perform two measurements, which we denote by $A, A'$, and $B, B'$ respectively. 
Let $\pr(X=a, Y=b)$ be the probability that Alice and Bob return answers $a$ and $b$ upon measuring $X\in\{A,A'\}$ and $Y\in\{B,B'\}$. 
The following conditions are required to hold. 
\begin{enumerate}
\item $\pr(A=0,B=0) > 0$, that is, for measurements $A$ and $B$ the outcome pair $A=0$ and $B=0$ is \emph{possible}, 
\item $\pr(A=0,B'=1) = 0$ and $\pr(A'=1,B=0)=0$, that is, for $A$ and $B'$ the outcome pair $A=0$ and $B'=1$ is \emph{impossible}, and similarly
for $A'=1$ and $B=0$,
\item $\pr(A'=0,B'=0)=0$, that is, for $A'$ and $B'$ the outcome pair $A'=0$ and $B'=0$ is \emph{impossible}.
\end{enumerate}
Hardy put forward quantum measurements and states which satisfy these conditions, and showed that no classical theory can be used to satisfy all three conditions. 
Quantumly, Alice and Bob measure their shared state $\ket{\psi}$ in one out of the two bases $\mathcal{S}$ or $\mathcal{S'}$, where the choice of the
unprimed basis will depend on the shared state. More precisely, we can pick any $0<\theta<\tfrac{\pi}{2}$ and consider a shared state
\be
  \ket{\psi_\theta} = \frac{1}{\sqrt{1+\cos^2(\theta)}}\Big(
  \sin(\theta) \ket{11} + 
  \cos(\theta) \Big( \ket{01} + \ket{10} \Big)
  \Big).
\label{eq:Shared}
\ee
Then we take $\mathcal{S}':=\{\ket{0},\ket{1}\}$ and  $\mathcal{S} := \{\ket{b_0},\ket{b_1}\}$, 
where
\begin{align}
  \ket{b_0} = & \sin(\theta) \ket{0} - \cos(\theta) \ket{1}\\
  \ket{b_1} = & \cos(\theta) \ket{0} + \sin(\theta) \ket{1}.
\end{align}
A simple calculation reveals that Conditions (1)--(3) are indeed satisfied.

A simple argument reveals that no local hidden variables theory can be compatible with Conditions (1)--(3).
When Alice and Bob follow the measurement procedure above on many states $\ket{\psi}$, then since $\pr(A=0,B=0)>0$ any local hidden variable theory will predict that $A=0$ \emph{and} $B=0$ for some of the states $\ket{\psi}$. Yet Condition~(2) implies that $A'=0$ and $B'=0$ for such particles, which violates Condition~(3).

\section{Unifying CHSH inequality and Hardy's paradox}
\label{sec:Unify}

At first glance, Bell inequalities and Hardy's paradox may seem fundamentally different beasts---Bell is concerned with expectation values and Hardy with the logical possibility and impossibilities of events. There have been several attempts to link Hardy's paradox to Bell inequalities and to understand its relation to the CHSH inequality in particular \cite{Mermin94, Garuccio95, Ghirardi08, Braun08}. As a continuation of these efforts, we unveil a close relationship between the CHSH inequality and Hardy's paradox that emerges when we are willing to view them through the goggles of non-local games.
In both scenario's we will assign a cost to a particular pair of 
measurement outcomes $a$ and $b$ given measurement settings $s$ and $t$, and the only differentiating factor will be the exact value of this cost. Indeed, one could consider a whole intermediary range of games that lie between Hardy's paradox and the CHSH inequality---each one assigning different costs. 

\paragraph{Non-local games as minimizing a cost function.}
\label{sec:Prelim}
To form this connection, let us now first explain the concept of a \emph{non-local game}.
The expert reader may thereby note that in the established literature, non-local games are sometimes expressed in terms of the probability $p_{\rm win}$ that Alice and Bob win a game. Here, we first propose a convenient reformulation by associating an overall cost function $c(G)$ any game $G$, where (up to normalization)
$p_{\rm win} = 1 - c(G)$. Mathematically, a two-party non-local game $G=(\mathcal{C},\pi)$ can be specified by a cost function $\mathcal{C}:  A\times B\times S\times T \to \R \cup \{\pm\infty\}$ and a probability distribution of inputs (measurement settings) $\pi: S\times T \to [0,1]$. The game starts when Alice receives input $s$ and Bob receives input $t$, where $(s,t)$ is chosen according to distribution $\pi$. Without communicating, the players output $a\in A$ and $b\in B$ respectively. The cost of their answers is then given by $\mathcal{C}(a,b|s,t):=\mathcal{C}(a,b,s,t)$. Before the start of the game players can agree on a strategy so to minimize the expected cost. For classical players we refer to the smallest achievable cost as the cost of the game and denote it by $c(G)$. In the case of quantum players we refer to the corresponding quantity as the quantum cost of the game and denote it by $c_q(G)$. 

In this note we only consider games with finite input and output sets. 
A classical strategy can be specified using two functions, $\alpha: S\times R \to A$ and $\beta: T\times R \to B$, where $R$ is the set of values taken by shared randomness. Upon input $s$ and shared randomness $r$ Alice outputs $\alpha(s,r)$; similarly upon input $t$ and shared randomness $r$ Bob outputs $\beta(t,r)$. If the shared randomness is distributed according  to $\tau:R\to[0,1]$, the cost of the game is given by
\be
  c(G) = \min_{\alpha,\beta} c(G|\alpha,\beta) =
   \min_{\alpha,\beta}
   \sum_{\substack{(s,t)\in S\times T\\ (a,b)\in A\times B} }
   \sum_{r\in R}
   \pi(s,t) \; \tau(r) \;
   \mathcal{C}\big(\alpha(s,r),\beta(t,r)|s,t\big),
\label{eq:CCost}
\ee
where $c(G|\alpha,\beta)$ is the cost incurred by strategy $(\alpha,\beta)$. Let us consider a strategy given by $\alpha_r(s):=\alpha(r,s)$ and $\beta_r(t):=\beta(r,t)$. Such strategies are said to be deterministic as they do not use any randomness. Noting that  
\be
  c(G|\alpha,\beta) = \sum_{r\in R} \tau(r)\; c(G|\alpha_r,\beta_r) 
\ee
shows that for some $r\in R$ we have  $c(G|\alpha,\beta) \ge  c(G|\alpha_r,\beta_r)$. 
Hence, the minimum in \Eq{CCost} is always achieved by some deterministic strategy and we can restrict our attention to such strategies only. Since these strategies do not make use of the randomness, we omit $R$ and minimize over $\alpha: S\to A$ and $\beta: T\to B$.

A quantum strategy can be specified by providing POVM's $\mathcal{A}^s=\{A_a^s : a\in A\}\subseteq\Pos(\C^{d_A})$ for each of Alice's input $s\in S$, POVMs $\mathcal{B}^s=\{B_b^t : b\in B\}\subseteq\Pos(\C^{d_A})$ for each of Bob's input $t\in T$ and a shared quantum state $\rho\in\Pos(\C^{d_A}\otimes\C^{d_B})$. Upon input $s$ Alice measures her part of $\rho$ using $\mathcal{A}^s$ and outputs the obtained measurement outcome. Similarly upon input $t$ Bob measures his part of $\rho$ using $\mathcal{B}^t$ and outputs the obtained measurement outcome. The quantum cost is given by
\be
  c_q(G) = \min \sum_{\substack{(s,t)\in S\times T\\ (a,b)\in A\times B} }
   \pi(s,t) \;  \mathcal{C}(a,b|s,t) \;
  \tr \Big(\big( A_a^s \otimes B_b^t \big) \rho \Big),
\label{eq:QCost}
\ee
where the minimization is performed over all quantum strategies. Any state $\rho$ is a probabilistic combination of some pure states and can thus be written as $\rho = \sum_i p_i \ket{\psi_i}\bra{\psi_i}$. Therefore, the cost incurred by a strategy that uses $\rho$ is a probabilistic combination of costs incurred by strategies using pure states $\ket{\psi_i}$. Thus, in \Eq{QCost} we can minimize only over strategies using pure states. We now cast both CHSH and Hardy's setup as non-local games by choosing appropriate cost functions.

\paragraph{CHSH game.}
\label{sec:CHSH}
It is well known that the CHSH inequality can be expressed as a non-local game~\cite{Brunner13}. Indeed our present formulation is merely a small twist in which we do not think about any probabilities that Alice and Bob win the CHSH game, but rather view their actions as trying to minimize an overall cost function.
For completeness sake, let us explain in detail on how we can 
express~\eqref{eq:chsh} in terms of a cost function. 
Note that since the outcomes of Alice and Bob's measurements are $\pm 1$, in order to maximize $\<CHSH\>$ Alice and Bob want to obtain different outcomes when measuring observable $A^1B^1$ and the same outcome in the remaining three cases. In other words, if  $s,t\in\{0,1\}$ are the measurement settings and $a,b\in\{+1,-1\}$ are the outcomes, then the answers $a$ and $b$ must satisfy $st=\abs{a-b}/2$. Here, we measure Alice's and Bob's success 
by a cost function that associates a unit cost with outcomes $a$ and $b$ that do \emph{not} satisfy $st=\abs{a-b}/2$; other outcomes incur no cost. 
What is the smallest cost that can be achieved classically and how does it compare to the cost achievable using quantum states and measurements?

Given any measurement settings $s$ and $t$, we have $\<A^s B^t\> = \pr(a=b|s,t) -\pr(a\neq b |s,t)$ and also $\pr(a=b|s,t) + \pr(a\neq b |s,t)=1$. So we can easily express the probability of obtaining the same (different) outcome in terms of expectations as 
\be
  \pr(a=b|s,t) = \frac{1+\<A^sB^t\>}{2} \;\text{ and }\; 
  \pr(a\neq b|s,t) = \frac{1-\<A^sB^t\>}{2}.
\ee
If we choose the measurement setting uniformly at random, the expected cost is
\begin{align}
  c(\text{CHSH}) &= \frac{1}{4}\Big( \pr(a\neq b|0,0)+\pr(a\neq b|0,1)
  +\pr(a \neq b|1,0)-\pr(a=b|1,1)\Big)\\
   &= \frac{1}{8}\Big(4- \<A^0 B^0\> - \<A^0 B^1\> 
  - \<A^1 B^0\> + \<A^1 B^1\> \Big)\\
  &= \frac{1}{2} \left( 1 - \frac{\<CHSH\>}{4}\right).
\end{align}
This shows that maximizing the CHSH expression from \Eq{chsh} is equivalent to minimizing a properly chosen cost function. Also, we get that the smallest classically achievable cost is $\tfrac{1}{2}(1-2/4) = 1/4$, while quantum mechanics allows to achieve cost as low as $\tfrac{1}{2}(1-2\sqrt{2}/4) = 
(2-\sqrt{2})/4 \approx 0.146$. 

As the above example shows, a non-local game is simply an alternative way of looking at Bell inequalities and their violations. 
In the non-local game setting, it is more convenient to express the cost incurred by specific quantum measurements in terms of 
outcome probabilities rather than expectation values of observables.
For example, suppose that upon challenges $s$ and $t$ Alice and Bob measure their part of shared state $\ket{\psi}$ using POVMs $\{A^s_0,A^s_1\}$ 
and $\{B^t_0,B^t_1\}$ where the subscript corresponds to the measurement outcome $a$ and $b$ respecitvely. 
We can now write the cost incurred by such a strategy as 
\be
  \frac{1}{4}\sum_{s,t} \sum_{a,b} \mathcal{C}(a,b|s,t)
  \pr(a,b|s,t)= 
  \frac{1}{4}\sum_{s,t} \sum_{a,b} \mathcal{C}(a,b|s,t)
  \bra{\psi} A^s_a \otimes B^t_b \ket{\psi},
\ee
where $\mathcal{C}$ is the cost function; for CHSH the cost function is $\mathcal{C}(a,b|s,t) = 1$ if $st\neq \abs{a-b}/2$ and $\mathcal{C}(a,b|s,t) = 0$ if otherwise. We refer the reader to \Sect{Unify} for a formal introduction to the non-local game framework.

Expressed in the language of games, quantum mechanics allows Alice and Bob to achieve a cost function that is strictly lower than what we can obtain by any classical means. As such, the cost yields a measure of how strong non-local correlations can be: a lower cost function means stronger non-local correlations.

Recall that in the CHSH setup the parties want to obtain the different outcomes on input $s=t=1$ and the same outcome otherwise. 
Let us now 
relabel the $\pm 1$ outcomes with binary values via $f(x) = \abs{1-x}/2$, \ie, we send $-1\to 1$ and $+1\to 0$. After this relabeling Alice and Bob want to ensure that their answers, $a\in\{0,1\}$ and $b\in\{0,1\}$ satisfy $s  t = a \oplus b$.
To encourage such outputs, we assign unit cost to answers that don't comply with the $s  t = a \oplus b$ requirement:
\newcommand{\smt}[4]{\begin{tabular}{cccc}#1&#2&#3&#4\end{tabular}}
\renewcommand{\arraystretch}{1.02}
\be
\begin{tabular}{l|c|c|c|c}
 & $s=t=0$ & $s=0,\; t=1$ & $s=1,\; t=0$ & $s=t=1$ 
\\ \hline
$a$ & \smt{0}{0}{1}{1} & \smt{0}{0}{1}{1} 
    & \smt{0}{0}{1}{1} & \smt{0}{0}{1}{1} 
\\ \hline
$b$ & \smt{0}{1}{0}{1} & \smt{0}{1}{0}{1} 
    & \smt{0}{1}{0}{1} & \smt{0}{1}{0}{1} 
\\ \hline\hline
$\mathcal{C}_{\text{CHSH}}(a,b|s,t)$ 
   & \smt{0}{1}{1}{0} & \smt{0}{1}{1}{0} 
   & \smt{0}{1}{1}{0} & \smt{1}{0}{0}{1} 
\end{tabular}
\label{eq:CHSH}
\ee
As explained in above the smallest expected classical cost is $c(\text{CHSH})=1/4$ while quantum mechanics allows us to reduce the cost to $c_q(\text{CHSH}) = (2-\sqrt{2})/4$

\paragraph{Hardy's paradox.}
\label{sec:Hardy}
Let us now assign different costs for outcome events from which we obtain Hardy's paradox. For the moment, let us consider the following assignment for the cost of outcome events. 
\be
\begin{tabular}{l|*{4}{c}|*{4}{c}|*{4}{c}|*{4}{c}}
   & \multicolumn{4}{c|}{$s=A,\; t=B$} 
   & \multicolumn{4}{c|}{$s=A,\; t=B'$}
   & \multicolumn{4}{c|}{$s=A',\; t=B$} 
   & \multicolumn{4}{c}{$s=A',\; t=B'$}
\\ \hline
$a$ & 0&0&1&1 & 0&0&1&1 & 0&0&1&1 & 0&0&1&1 
\\ \hline
$b$ & 0&1&0&1 & 0&1&0&1 & 0&1&0&1 & 0&1&0&1 
\\ \hline\hline
$\mathcal{C}_{\text{Hardy}}(a,b|s,t)$ 
   & 0        & $T$      & $T$      & $T$
   & 0        & $\infty$ & 0        & 0 
   & 0        & 0        & $\infty$ & 0 
   & $\infty$ & 0        & 0        & 0
\end{tabular}
\label{eq:Hardy}
\ee
To strictly enforce Conditions (2) and (3) we have assigned infinite cost for outputting $(a,b)=(0,1)$ upon input $(A,B')$ or outputting $(a,b)=(1,0)$ upon input $(A',B)$ or outputting $(a,b)=(0,0)$ upon input $(A',B')$. To encourage the players to output $(a,b)=(0,0)$ upon input $(A',B')$ we assign any nonzero cost $T$ to the other three possible outputs. This definition of cost function allows us to view Hardy's paradox as a non-local game $G_{\text{Hardy}} = (\mathcal{C}_{\text{Hardy}}, \pi)$, where we take $\pi$ to be the uniform distribution. 

Note that $c_q(G_{\text{Hardy}})<\infty$, since quantum players can use shared state $\ket{\psi_\theta}$ and perform the primed and unprimed measurements as described above to satisfy Conditions~(2) and (3) thus avoiding infinite cost. To further minimize the cost they can choose $\theta$ in \Eq{Shared} so to maximize the probability that both Alice and Bob output zero upon unprimed inputs. It can be verified that the maximum is $(5\sqrt{5}-11)/2\approx0.09$ and is achieved at $\theta=\arccos\big(\big(\tfrac{\sqrt{5}-1}{2}\big)^{1/2}\big) < \frac{\pi}{4}$. Hence, 
\be
  c_q(G_{\text{Hardy}}) \le 
  \frac{T}{4} \Big( 1-\frac{5\sqrt{5}-11}{2} \Big) 
  \approx 0.23\;T.
\ee

Let us now argue that the classical cost of $G_\text{Hardy}$ exceeds the quantum one. The argument is essentially the same as the one used to show that no local hidden variables theory can be used to explain observations compatible with Conditions (1)--(3). As explained above, we can restrict our attention to deterministic strategies, \ie, a pair of functions $\alpha:\{A,A'\} \to \{0,1\}$ and $\beta:\{B,B'\} \to \{0,1\}$. Note that if Alice and Bob want 
to incur cost strictly smaller than $T/4 > c_q(G_{\text{Hardy}})$, they need to set $\alpha(A) = 0$ and $\beta(B) = 0$. After having fixed this choice, to avoid incurring infinite cost on inputs $(A,B')$ and $(B,A')$ they need to set $\alpha(A')=0$ and $\beta(B')=0$. Yet this leads to infinite cost on input $(A',B')$. The only deterministic strategy that incurs finite cost is for both Alice and Bob to always output 1 irrespective of the input. Hence, $c(G_\text{Hardy}) = T/4$.

Of course, it is intuitive that assigning an infinite cost to an event encourages Alice and Bob to make sure that this event is in fact impossible, which is the essence of Hardy's paradox. Nevertheless, it is interesting to note that in this reformulation it was not necessary to assign an infinite cost to some events in order to obtain the same non-local behavior. In fact, we could replace the infinite cost in Table~\ref{eq:Hardy} with any sufficiently large value (for example, any value larger than $T$ would do). This way, any optimal strategy would have to avoid the outputs forbidden in Hardy's paradox and the optimal expected cost, both classical and quantum, would remain 
identical.

\paragraph{Parametrized family of games.}
\label{sec:Param}

Looking the above, it is clear that Hardy's paradox and the CHSH inequality are not the only two measures of non-locality that we could obtain
in this fashion. In particular, we could define a two parameter family of two-input two-output games. We will show that both Hardy's paradox and CHSH game can be recovered by choosing appropriate values of the parameters.

We consider a family of games $G(\phi,w) = (\pi, \mathcal{C}_{\phi,w})$ parametrized by for $\phi\in[0,\pi/2]$, $w\in\R^+$. We fix $\pi$ to be the uniform distribution and define the cost function as follows:
\be
\begin{tabular}{l|*{4}{c}|*{4}{c}|*{4}{c}|*{4}{c}}
   & \multicolumn{4}{c|}{$s=t=0$} & \multicolumn{4}{c|}{$s=0,\; t=1$}
   & \multicolumn{4}{c|}{$s=1,\; t=0$} & \multicolumn{4}{c}{$s=t=1$}
\\ \hline
$a$ & 0&0&1&1 & 0&0&1&1 & 0&0&1&1 & 0&0&1&1 
\\ \hline
$b$ & 0&1&0&1 & 0&1&0&1 & 0&1&0&1 & 0&1&0&1 
\\ \hline \hline 
$\mathcal{C}_{\phi,w}$ 
   & 0       & $\cos\phi$ & $\cos\phi$ & $\sin\phi$
   & $0$     &  $1/w$     & $w$        & 0 
   & $0$     &  $w$     & $1/w$        & 0 
   & $1/w$   &  0         & 0          & $w$ 
\\ \hline\hline
$\mathcal{C}_\text{Hardy}$
   & 0        & $T$      & $T$      & $T$
   & 0        & $\infty$ & 0        & 0 
   & 0        & 0        & $\infty$ & 0 
   & $\infty$ & 0        & 0        & 0
\\  \hline
$\mathcal{C}_\text{CHSH}$ 
& 0&1&1&0 & 0&1&1&0 & 0&1&1&0 & 1&0&0&1
\end{tabular}
\label{eq:Family}
\ee
For comparison, we have included CHSH and Hardy's cost functions in the last two rows. We have also relabeled the unprimed inputs in Hardy's paradox as `0' and the primed inputs as `1'.
From the above table one can easily see that
setting $\phi:=0$ and $w:=1$ gives the CHSH cost function. Hence, $G(0,1)$ gives us the CHSH game. Similarly, by setting $\phi:=\pi/4$, $w:=0$ in Table~\ref{eq:Family} and following the convention $1/0 = \infty$ we recover the Hardy's cost function. Hence, $G(\pi/4,0)$ corresponds to Hardy's paradox.

\section{Conclusion}

We have shown that the CHSH inequality and Hardy's paradox can be understood as two special cases of a much more general family of non-local 
games, or equivalently, Bell inequalities. Considering the fifty year anniversary of Bell inequalities it is perhaps satisfying that we can understand the fundamental contributions of both Bell and Hardy under one umbrella. In addition to being conceptually pleasing, this opens up a possibility to look for new ``Hardy-like'' paradoxes by viewing existing Bell inequalities as games with a cost function and increasing the cost associated with some of the questions.


We also note that our reformulation in terms of the cost function would allow a clear statement for experimental verifications~\cite{hardy1,hardy2, hardy3} 
of Hardy's paradox in which the event labelled ``never'' in Figure~\ref{fig:Hardy} is virtually certain to occur eventually due to experimental 
imperfections, prompting workarounds~\cite{Mermin94,Braun08}. 
As such, assigning it a 0 probability in the estimation is certainly inaccurate. What's more, even if this event was never observed, no finite experiment 
can completely rule out its existence. Our framework, however, allows these situations to be dealt with in a quantifiable 
way by choosing a large, but finite, $T < \infty$ in the cost function. That is, the maximum value of $T$ admissible to see a quantum violation 
of the non-local game corresponding to Hardy's paradox yields a natural measure of the confidence of the test.

It is clear that we still do not fully grasp the extent of quantum non-locality, its limitations and its consequences. In particular, even given considerable efforts it remains difficult to explain why quantum mechanics itself is not 
more non-local, that is, why is quantum non-locality---or the smallest possible cost attainable---not merely limited by the non-signalling principle~\cite{PR1,PR2,PR3}? Much work has gone into understanding such limitations, especially for the CHSH inequality itself, by imagining a world that exhibits 
more non-locality~\cite{PPKS+09,vandam,ow10,bbb+09}. It would be interesting to understand what the implications of stronger than quantum non-locality 
would be for the above class of games, containing in particular Hardy's inequality as a special case. 
Another future research direction that naturally arises from the presented unification is the investigation of games $G(\phi,\omega)$ for other values of the parameters $\phi$ and $\omega$. For instance, do certain values of the parameters give rise to already known Bell inequalities or non-local games? Also, since the optimal state for the CHSH inequality differs from that of Hardy's paradox, one could examine how it changes as we interpolate the parameters between the values giving these two scenarios.

\paragraph{Acknowledgments.}
This research is supported by the Ministry of Education (MOE) and National Research Foundation Singapore, as well as MOE Tier 3 Grant ``Random numbers from quantum processes'' (MOE2012-T3-1-009).

\bibliographystyle{alphaurl}
\newcommand{\etalchar}[1]{$^{#1}$}


\end{document}